\documentclass[aps,prl,superscriptaddress,showpacs,twocolumn,floatfix]{revtex4}
\usepackage{bm}
\usepackage{graphicx}
\usepackage{epsfig}
\usepackage{color}

\begin{document}

\newcommand{\qav}[1]{\left\langle #1 \right\rangle}
\newcommand{\myT}{\Gamma}
\newcommand{\rem}[1]{}
\newcommand{\refe}[1]{(\ref{#1})}
\newcommand{\fige}[1]{Fig.~\ref{#1}}
\newcommand{\refE}[1]{Eq.~(\ref{#1})}
\newcommand{\beq}{\begin{equation}}
\newcommand{\eeq}{\end{equation}}
\newcommand{\beqa}{\begin{eqnarray}}
\newcommand{\eeqa}{\end{eqnarray}}
\newcommand{\cg}{\check g}
\newcommand{\inc}{{\rm inc}}
\newcommand{\Pc}{{\cal P}}
\newcommand{\larghezza}{7.8cm}

\title{The charge shuttle as a nanomechanical ratchet}
\author{F. Pistolesi}
\affiliation{Laboratoire de Physique et Mod\'elisation des Milieux
    Condens\'es, CNRS-UJF B.P. 166, F-38042 Grenoble, France}
\author{Rosario Fazio}
\affiliation{ NEST-INFM $\&$ Scuola Normale Superiore
    Piazza dei Cavalieri 7, Pisa Italy}
\date{\today}

\begin{abstract}
We consider the charge shuttle proposed by Gorelik {\em et al.}
driven by a time-dependent voltage bias.
In the case of asymmetric setup, the system behaves as a rachet.
For pure AC drive, the rectified current shows a complex
frequency dependent response characterized by frequency locking at
fracional values of the external frequency.  Due to the non-linear
dynamics of the shuttle, the rachet effect is present also for very
low frequencies.
\end{abstract}

\maketitle

The great burst in the study of Nano-Electro-Mechanical (NEMS)
devices is unvealing several new perspectives in the
realization of nanostructures where the charge transport is assisted
by the mechanical degrees of freedom of the device itself.
This is the case, for example, when nanomechanics~\cite{cleland}
has been combined with single electron tunneling.
Important experiments in this area are the use of a
Single Electron Transistor (SET) as
dispacement sensor~\cite{knobel} or quantum transport through
suspended nanotubes~\cite{suspended}, oscillating
molecules~\cite{park,smit,kubatkin,pasupathy} and
islands~\cite{erbe}.
On the theoretical side, several
works~\cite{gorelik,weiss,boese,braig,sapmaz,novotny,fedorets,
armour,pistolesi,novotny2,chtchelkatchev,blanter}
have already highlighted various aspects of the role of mechanical
motion on single electron tunneling.

An exciting prototype example of mechanical assisted SET device has
been proposed by Gorelik {\em et al.}~\cite{gorelik} and named Single
Electron Shuttle.
The authors of Ref.~\onlinecite{gorelik} predicted that a SET with an
oscillating central island can shuttle electrons between the
electrodes leading to low noise transport~\cite{weiss,pistolesi,novotny2}.
Although the realization of the charge shuttle is difficult
experimentally, promising systems are $C^{60}$ molecules in break
junctions~\cite{park,pasupathy}, or silicon structures~\cite{erbe,blick}.

One of the advantage of this self-oscillating structure is the fact
that one can generate very high frequency mechanical oscillation with
static voltages.
The fact that the system has an intrinsic and stable oscillating mode
as the result of a static voltage suggests that the application of an
oscillating voltage may lead to new interesting effects, related to the
interplay between the external AC drive and the internal frequency of
the device.
Moreover, as the non-linearities of the dynamics have an important role, this
interplay should emerge in a wide interval of the ratio of the two
frequencies.
Aim of this Letter is to study a shuttle driven by a
time-dependent applied bias.
The most interesting situation is when the system is
asymmetric.
We will show that the structure acts as a ratchet~\cite{ratchet} in
which the forcing potential is generated in a self-consistent way.
We study in some details the case in which the external bias is AC and
report a quite rich behaviour as a function of applied frequency.
We find clear indications of frequency locking.
The resulting DC current may have both signs, depending on the value
of the frequency.
The response to an AC field can give strong
indications on the motion of the central island, even when the system
is very far from the shuttling instability.
A sizeable ratchet effect is present down to
frequencies much smaller than the mechanical resonating frequency, due
to the adiabatic change of the equilibrium position of the grain.
In a very recent experiment, Scheible and Blick~\cite{blick},
already observed similar results to those presented in our
work.

%
%
\begin{figure}
    \centerline{\epsfig{file=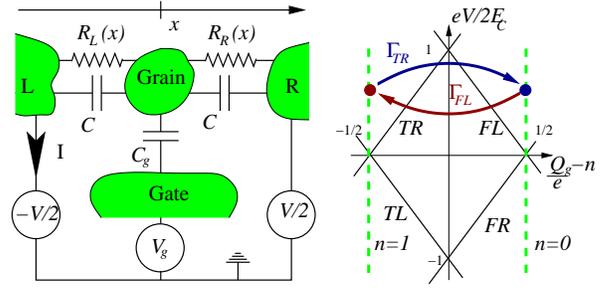,width=\larghezza}}
    \caption{Left panel: Schematic of a charge shuttle.
    Right panel: energy diagram for the SET. The diagonal lines indicate the
    thresholds for the vanishing of the four rates $\Gamma_{FL}$,
    $\Gamma_{FR}$,  $\Gamma_{TL}$, and  $\Gamma_{TR}$.
    The two dots indicate the state of the system during shuttling at
    fixed voltage $V$.}
\label{fig1}
\end{figure}
%
%

The single electron shuttle, shown in \fige{fig1}, is a
SET where the central island can oscillate between the two
leads~\cite{gorelik}.
The central island is subjected to an elastic recoil force, a damping
force due to the dissipative medium, and an electric force due to the
applied bias.
The island is connected to the left and right leads through tunnel
junctions with resistances $R_L(x)=R_L(0) e^{x/\lambda}$ and
$R_R(x)=R_R(0) e^{-x/\lambda}$, ($\lambda$ is the tunnelling length and
$x$ is the displacement from the equilibrium position in absence of
any external drive).
The capacitances $C_g$ and $C$ couple the island to the gate and to
the two leads, respectively.
We neglect small variations of the capacitances due to the motion of
the island as this dependence is weak when compared with the
exponential dependence of the tunneling resistances on $x$.
The system is biased symmetrically at a voltage $V(t)$
($V_R=-V_L=V/2$), the charging energy is $E_C=e^2/2C_\Sigma$ where
$e$ is the electron charge, $C_\Sigma=C_g+2C$, and the gate charge $Q_g$
is $C_g V_g$.
For simplicity, we consider the case of low temperatures ($T \ll E_C$),
charge degeneracy ($Q_g = e/2$), and voltages $|V|<E_C/e$.
In this regime the grain can accommodate only $n=0$ or $1$ additional
electrons (see right panel of \fige{fig1}).
All the properties of the nanomechanical ratchet are captured already
at this level.

In the simplest approximation the dynamics of the central island is
described by Newton's law~\cite{gorelik}
\beq
\ddot x(t) = -\omega_o^2\, x(t) - \gamma\, \dot x(t) + {eV(t)\over mL}\, n(t)
\;\; .
        \label{xeq}
\eeq
Here $m$ is the mass of the grain, $\omega_o$ is the oscillator
eigenfrequency, $\gamma$ is a damping coefficient and $L$ is the
distance between the two leads.
In the regime of incoherent transport the (stochastic) evolution of
the charge $-en(t)$ is governed by the following four
rates~\cite{averin91}:
$\Gamma_{FL} = |eV(t)/4E_C| \Gamma_L(x) \Theta(V)$
and
$\Gamma_{FR} =  |eV(t)/4E_C| \Gamma_R(x) \Theta(-V)$
for $n=0 \rightarrow 1$ transitions,
$\Gamma_{TL}=  |eV(t)/4E_C| \Gamma_L(x)\Theta(-V)$
and $\Gamma_{TR} = |eV(t)/4E_C| \Gamma_R(x) \Theta(V)$
for $n=1\rightarrow0$ transitions.
Here $FL$, $FR$, $TL$, and $TR$ stands for From/To and Left/Right,
indicating the direction for the electron tunneling associated to the
corresponding,
$\Gamma_{L/R}(x)=[R_{L/R}(x)C]^{-1}$ and $\Theta(t)$ is the Heaviside
function.
The current $I$ is then determinated by counting
the net number of electrons that have passed through the system in a
given time $t$.

\begin{figure}
\centerline{\psfig{file=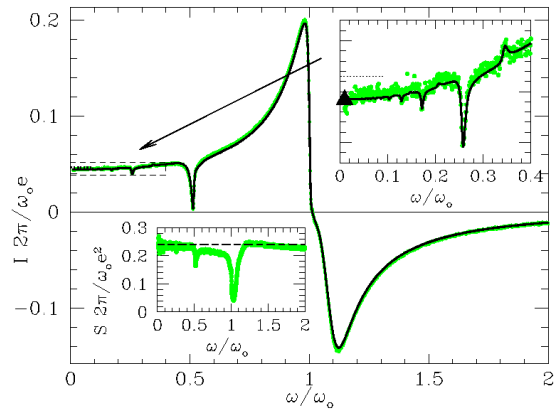,angle=-90,width=\larghezza}}
\caption{Current as a function of the frequency for $\epsilon=0.5$,
  $\gamma/\omega_o=0.05$, $\Gamma/\omega_o=1$, and $R_R/R_L=10$. The result of the
  simulation of the stochastic dynamics (points) is compared with the
  approximate $I_a$ (full line). In the small frequency region, enlarged
  in the inset, several resonances at fractional values of $\omega_o$ appear.
  We also show (dotted) the analytical result from~\refE{adiabatic}
  in the adiabatic limit. The triangular dot indicates the numerical solution
  of the adiabatic equations.
  Lower inset: current noise  from the simulation (points) and
  analytic result (dashed) for the static SET.
}
\label{fig2}
\end{figure}

In most of the paper we consider the case of an oscillating voltage bias
$V(t)=V_o \sin(\omega t)$~\cite{pistolesi2}, where the interplay of
the two frequencies $\omega$ and $\omega_o$ is crucial.
If the system is perfectly symmetric no direct current can be
generated, since this would break the left/right symmetry.
We then concentrate on the asymmetric case $R_R(0)/R_L(0) \ne 1$.
We performed simulation of the stochastic process governed by the four
rates defined above and by \refE{xeq}.
The results presented are obtained by simulating
$10^6$ events of tunneling for each plotted point.
After a transient time the system reaches a stationary behavior.
In \fige{fig2} we show the stationary DC current as a function of
the frequency of the external bias.
The rich structure shown in the figure is generic, we observed a
qualitatively identical behaviour in a wide range of parameters.
The existence of a direct current as a result of an applied periodic
modulation shows that the charge shuttle, whose stochastic dynamics
has been defined above, behaves as a ratchet~\cite{ratchet}.
Since our system is non linear, the external driving  will affect the
dynamics also for values of $\omega$ very different from the
natural frequency $\omega_o$.
Note that in this model the non-linearities are intrinsic to the shuttle
mechanism. They are not due to a non-linear mechanical force,
but they stems from the time dependence of $n(t)$.
As it is evident from \fige{fig2} the ratchet behaviour is present
also in the adiabatic limit $\omega/\omega_o \ll 1$.
In addition a series of resonances, due to frequency locking~\cite{CaosBook}
when $\omega \approx \omega_o \,q/p$, with $q$ and $p$
integers.
In this case the motion of the shuttle and the oscillating
source become synchronized in such a way that every $q$ periods of
the oscillating field the shuttle performs $p$ oscillations.
In \fige{fig2} we also report the low frequency current noise (lower
inset).
The net DC current results from large cancellations between positive
and negative contributions.
But the current noise is always positive, thus it does not cancel.
We find that the noise remains very close to the value for a SET
$S=e^2 \Gamma_L\Gamma_R(\Gamma_L^2+\Gamma_R)^2/(\Gamma_L+\Gamma_R)^3$
shown with a dashed line in  \fige{fig2}.
At resonance, more ordered transport is realized and
current becomes less noisy \cite{pistolesi}.
In the following we  provide a more detailed analysis by solving the
problem analytically in some tractable limits and by analyzing in more
details the behaviour of the shuttle in the frequency locked case.
%

%
%
%
%

If the electric force is much smaller than the mechanical one,
$\epsilon = e V_o/(\omega_o^2 m \lambda L)\ll 1$,
one can take into account the force generated by the stochastic
variable $n(t)$ only on average by substituting into \refE{xeq} its
mean: $\qav{n(t)}=P(t)$ where $P(t)$ is the probability to have
occupation $n=1$ in the grain.
The charge dynamics in the central island is then described by a
simple master equation~\cite{averin91}
\beq
     \dot P(t)
     = -\Gamma_1(t) P(t) + \Gamma_2(t)
     \,,
     \label{Peq}
\eeq
with
$
    \Gamma_1(t)= |eV(t)/4E_C| \left(\Gamma_L[x(t]+\Gamma_R[x(t)]\right)
$ and
$
    \Gamma_2(t)=
    |eV(t)/4E_C| \left(
    \Gamma_L[x(t]\Theta[v(t)]+\Gamma_R[x(t)]\Theta[v(t)]
    \right)
$.
The rates depend on the time $t$ through the voltage $V(t)$ and the
position of the grain $x(t)$.
The instantaneous (average) current through the structure is
\beq
    I_a(t)/e
    =
    \left(1-P(t)\right)\Gamma_{FL}-P(t) \Gamma_{TL}(t)
    \quad,
    \label{Ileft}
\eeq
where the subscript in the current indicates that the fluctuations of
the force acting on the shuttle, due to the discrete nature of the charge
tunneling, are neglected.
As shown in Ref.~\onlinecite{gorelik} the shuttle instability, at constant
bias, is controlled by the ratio $\epsilon\, \omega_o/\gamma$, we can
thus assume both parameters small, but their ratio arbitrary.

In the adiabatic limit
($\omega \ll \omega_o$ and  $\epsilon \,\omega_o/\gamma \ll 1$)
it is possible to find an approximate solution of
Eqs.~\refe{xeq} and \refe{Peq}.
In this case the position of the grain is given by the local
stationary solution of \refE{xeq}
$ x/\lambda = \epsilon \Gamma_2(x,V)/ \Gamma_1(x,V)$.
Solving this equation in lowest order in $\epsilon$ one obtains
\beq
    I_{a}(\omega \ll \omega _o) =
    \epsilon \, { V_o^2  e^3 \over 32 E_C^2}
    {\Gamma_L \Gamma_R
        (\Gamma_L-\Gamma_R) \over (\Gamma_R+\Gamma_L)^2}
        \,.
    \label{adiabatic}
\eeq
The corresponding value is shown in \fige{fig2}.
The (small) difference with the full numerics is due to the
expansion in $\epsilon$.
By solving numerically the equation for the local equilibrium position of
the grain, one obtains the result indicated by the triangular dot
at $\omega=0$ in the inset of \fige{fig2}~\cite{footnote}.
If $\epsilon\,\gamma/\omega_o$ is larger than the critical value for shuttling,
the behavior is completely different.
The grain will oscillate at its natural frequency $\omega_o$ with the
amplitude slowly modulated by $V(t)$.
The modulation will be small, since for small $\epsilon$ the effect of
a change in the value of $V$ affects the mechanical motion only after
several oscillations.
In this case the rectified current is much smaller than in the adiabatic
limit.

We now discuss in more details the dependence of the current on the
external frequency.
The most prominent structure, observed also in the experiments of
Ref.~\onlinecite{blick}, is present in our simulations at $\omega\approx
\omega_o$, and it corresponds to the main mechanical resonance.
The current changes sign accross the resonance.
This behaviour is due to the phase relation between the driving
voltage and the dispacement of the grain.
We verified this conjecture by solving
\refE{Peq} with $x(t)= A \sin(\omega t-\phi)$.
By calculating the current as a function of $\phi$ and using the
usual resonant dependence for
$\phi(\omega) = \arctan[(\omega^2-\omega_o^2)/\omega\eta ]$
we could reproduce qualitatively the behavior of \fige{fig2}.

\begin{figure}
    \centerline{\psfig{file=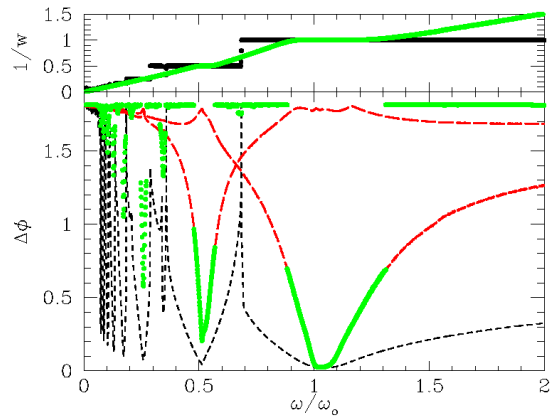,angle=-90,width=\larghezza}}
    \caption{
    The top panel
    shows the winding number obtained from the stochastic simulation
    (light points) and from the average approximation (black points).
    In the bottom panel $\Delta \phi$ is obtained from the average
    approximation (dashed line), from the stochastic simulation using
    $w$ from \refE{wdef} (light points). The variance has also been
    calculated from the simulation in the case of $w=1$ and $w=2$
    (long dashed line).
    The parameters are the same as in \fige{fig2}
}
\label{fig3}
\end{figure}

\begin{figure}
\centerline{\psfig{file=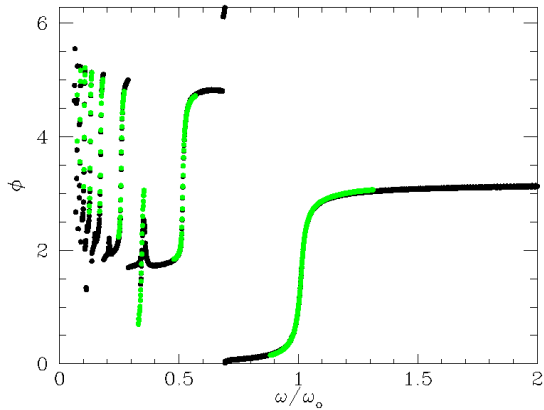,angle=-90,width=\larghezza}}
\caption{
Phase shift $\phi$ (same parameters and same notation of \fige{fig2}).
The phase shift is shown only when $\Delta \phi(\omega) < 1.7$
(cfr. \fige{fig3}).
When locking is achieved in presence of the stochastic
fluctuations $\phi$ remains the same of the
average simulation.}
\label{fig4}
\end{figure}

From \fige{fig2} it is clear that additional structures appear also
for $\omega \approx \omega_o/q$ (magnified in the upper inset),
with $q=2, 3, \dots $ (the numerical results indicate that, except for
the fundamental frequency, even $q$ are favorite with respect to odd
ones).
As we already anticipated, the motion of the shuttle and the
oscillating source become synchronized at commensurate frequencies
whose ratio is $p/q$.
This ratio, known also as the winding number,
can be defined more precisely as
\beq
   w =
   \lim_{t\rightarrow \infty}
   {\theta(t)/\omega t}
    \label{wdef}
\eeq
where $\theta(t)$ is the accumulated angle of rotation of
the representative point $(\dot x,\ddot x)$
[$\theta(t)/2\pi$ gives the number of oscillations
performed by the shuttle during the time $t$].
When the system is frequency locked at a winding number $w$, it is
possible to define the phase shift $\phi(t) = \theta(t)-w \, \omega
t$ \cite{nota2}.
After a transient time for perfect locking $\phi$ should not
depend on $t$  (apart from a small fluctuation if the motion is not
perfectly harmonical).
An additional important quantity to analyze is thus the phase shift variance
$ (\Delta \phi)^2 = \qav{\phi^2}-\qav{\phi}^2$.
This is calculated by sampling $20$ points per cycle over $10^3$ cycles.
For a given $w$ the smaller is the value of $\Delta\phi$, the better the system
locks to that external frequency.
The numerical results for $w$, $\phi$, and $\Delta \phi$ are
shown in \fige{fig3} and \fige{fig4}.
%
%
%
%

In \fige{fig3} we show the dependence of the
winding number as a function of the external frequency (top panel)
together with the analysis of $\Delta \phi$ (lower panel)
calculated from the stochastic simulation and from the average
approximation (short dashed line).
The locking at rational winding numbers is confirmed by the presence of
plateaux of decreasing width.
As expected the most stable plateau is at $w=1$: the system locks
very well at this frequency.
We find that this holds up to very high frequency in the
average approximation, where $w=1$ seems the only possible winding number.
The stochastic simulation would indicate instead that for
$\omega/\omega_o \gtrsim  1.3$
the locking with $w=1$, as defined from \refE{wdef}, is no more established.
However, by studying $\Delta \phi$ for $w=1$ in the whole frequency
range, one actually obtains that a correlation is always present
({\em i.e.} $\Delta\phi < \pi/\sqrt{3}\approx 1.81$)
even when \refE{wdef} gives a value of $w$ different from one.
The stochastic fluctuations thus unlock the shuttle globally, but not locally.
Looking at the presence of local locking at other winding numbers
we find that for instance $w=2$ is clearly present for
$\omega>1$ and reversely $w=1$ is present around $\omega=1/2$
(see long dashed lines in lower panel of \fige{fig3}).
For global locking, only one phase variance is minimal.
It corresponds to the ``dominant'' winding number.
The presence of correlations of other winding numbers
may indicate partial locking at these winding numbers
(as in the region $0.6 \lesssim \omega \lesssim 0.9 $ for $w=1$ and $2$)
or the contribution of higher armonics of $x(t)$
(as in the region $\omega>1$).

%
%
%
%

The dependence of the phase shift, see \fige{fig4}, around each
resonance is also remarkable. It is very similar to the behavior of a
forced harmonic oscillator, but instead of evolving from 0 to $\pi$ it goes
from $\phi_o$ to $\phi_o+\pi$, where $\phi_o$ depends on the
resonance.
With the aim of understanding the behaviour of the current close to the
resonance, we also considered the adiabatic limit, and looked for harmonic
oscillations superimposed to the adiabatic solution given before.
This can be verified numerically by searching a periodic solution of
period $2\pi/\omega$ for $P(t)$.
We find that in this case the direct current is non vanishing and that
it depends strongly on $\phi$.
Using the phase dependence given by the full numerical calculation we
could reproduce the shapes of the resonances.

In this Letter we showed that the charge shuttle under time-dependent
driving behaves as a ratchet. In the case of a AC bias the response
of the shuttle is rather rich due to the non-linear dynamics of the
grain. All the results obtained here can be tested experimentally.
indeed the very recent paper by Scheible and Blick~\cite{blick}
already reports on some of the properties related to the main
resonance.

We thank F. Faure and A. Romito for very useful discussions.
We acknowledge financial support from CNRS through contract
ATIP-JC 2002 (F.P.) and EU-RTN-Nanoscale, PRIN-2002,
Firb (R.F.).


\begin{thebibliography}{}
\bibitem{footnote}
    The racthet effect can be also detected in a simple way by
    injecting low frequency noise in the ystem.
    The resulting current, in the low frequency regime,
        $I_{a} =
        \epsilon\,(e^3 P_V /4 E_C^2) \Gamma_L \Gamma_R
        (\Gamma_L-\Gamma_R)/ (\Gamma_R+\Gamma_L)^2
        $
    is proportional to the integrated power spectrum of the noise
    $P_V=\langle V^2\rangle$.
\bibitem{nota2}
        For an accurate analysis, the value used for $w$ in the definition
        of $\phi(t)$ must be the best $p/q$ approximation to $w$, since even
    corrections of the order of $10^{-3}$ to the exact $p/q$ results
    would accumulate in the calculation of $\phi$, and give an arbitrary
    shift of the order of $\pi$.
\end{thebibliography}
\end{document}